\documentclass[runningheads]{llncs}

\usepackage{graphicx}
\usepackage{hyperref}
\usepackage{booktabs}
\usepackage{microtype}
\usepackage{numprint}

\usepackage{xcolor}

\begin{document}

\title{Synthetic data for unsupervised polyp segmentation \thanks{This project has received funding from the European Union’s Horizon 2020 research and innovation programme under the Marie Skłodowska-Curie grant agreement No 765140. This publication has emanated from research supported by Science Foundation Ireland (SFI) under Grant Number SFI/12/RC/2289\_P2, co-funded by the European Regional Development Fund.}}

% \titlerunning{Synthetic data for unsupervised polyp segmentation}
\author{Enric Moreu\inst{1,2}\orcidID{0000-0003-1336-6477} \and
Kevin McGuinness\inst{1,2}\orcidID{0000-0003-1336-6477} \and
Noel E. O'Connor\inst{1,2}\orcidID{0000-0002-4033-9135}}

\authorrunning{Enric Moreu et al.}

\institute{Insight SFI Centre for Data Analytics, Ireland \and
Dublin City University, Ireland
}

\maketitle 
\begin{abstract}
Deep learning has shown excellent performance in analysing medical images. However, datasets are difficult to obtain due privacy issues, standardization problems, and lack of annotations. We address these problems by producing realistic synthetic images using a combination of 3D technologies and generative adversarial networks. We use zero annotations from medical professionals in our pipeline.
Our fully unsupervised method achieves promising results on five real polyp segmentation datasets.
As a part of this study we release Synth-Colon, an entirely synthetic dataset that includes \numprint{20000} realistic colon images and additional details about depth and 3D geometry: \url{ https://enric1994.github.io/synth-colon}

\end{abstract}

\section{Introduction}

Colorectal cancer is one of the most commonly diagnosed cancer types. It can be treated with an early intervention, which consists of detecting and removing polyps in the colon. The accuracy of the procedure strongly depends on the medical professionals experience and hand-eye coordination during the procedure, which can last up to 60 minutes. Computer vision can provide real-time support for doctors to ensure a reliable examination by double-checking all the tissues during the colonoscopy.

\begin{figure}
\centering
\includegraphics[width=60px]{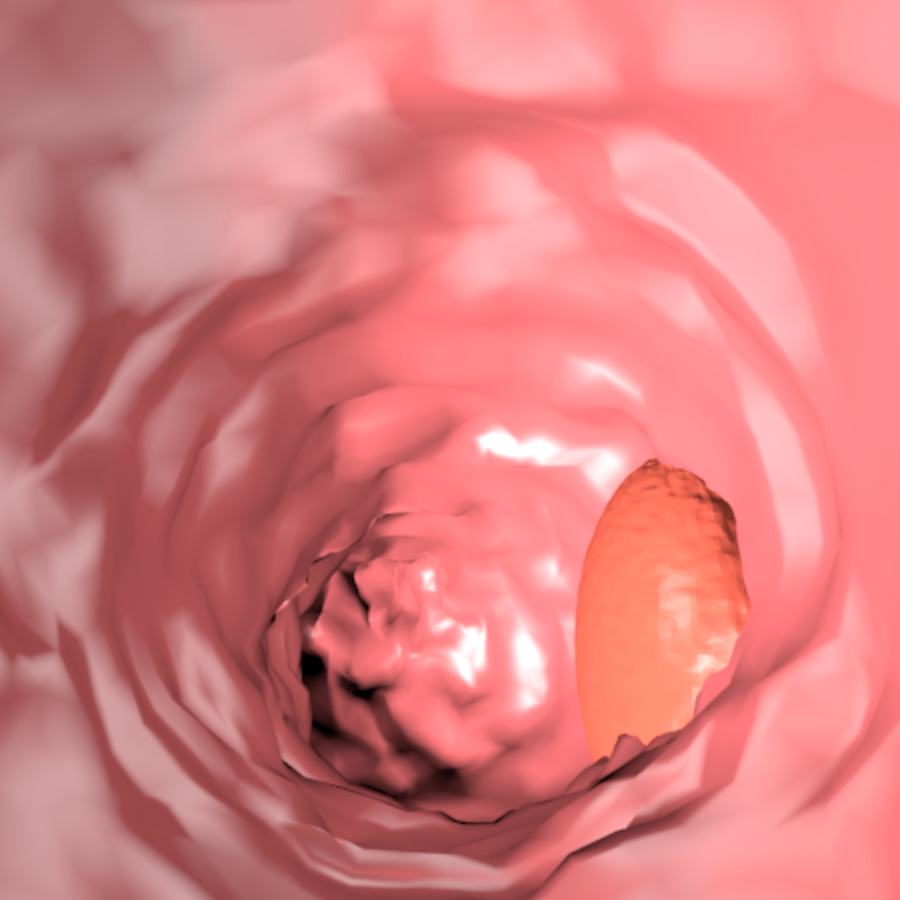}
\includegraphics[width=60px]{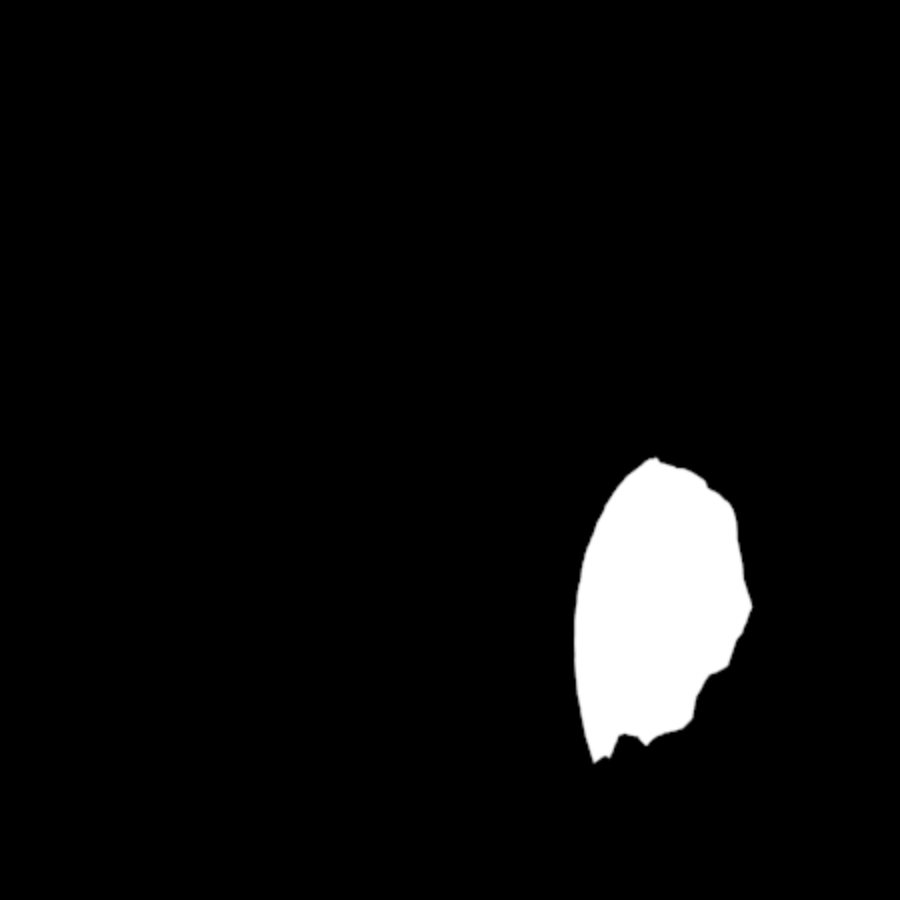}
\includegraphics[width=60px]{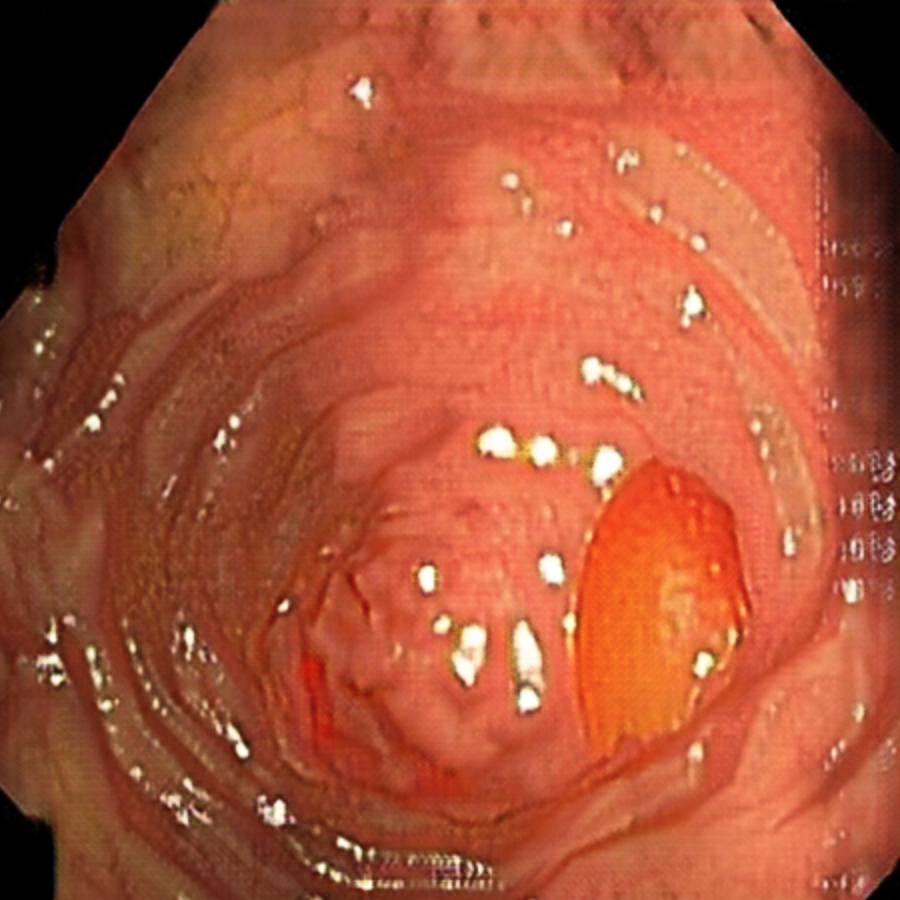}
\includegraphics[width=60px]{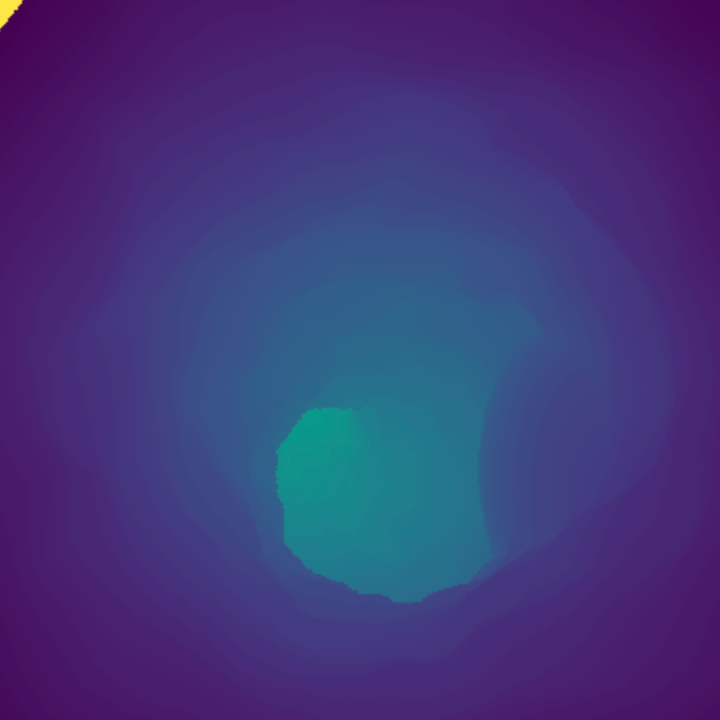}
\includegraphics[width=60px]{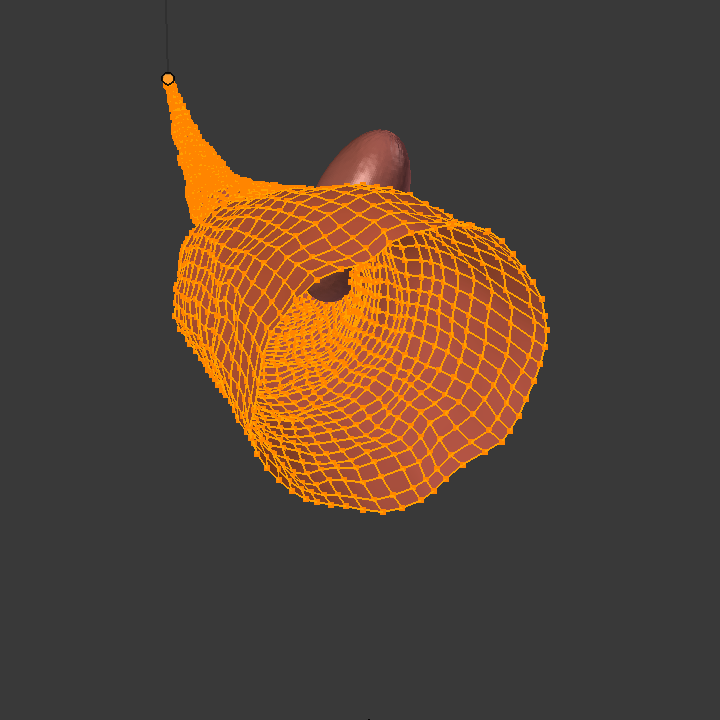}
\caption{Synth-Colon dataset samples include: synthetic image, annotation, realistic image, depth map, and 3D mesh (from left to right).} \label{dataset}
\end{figure}

The data obtained during a colonoscopy is accompanied by a set of issues that prevent creating datasets for computer vision applications.
First, there are privacy issues because it is considered personal data that can not be used without the consent of the patients. 
Second, there are a wide range of cameras and lights used to perform colonoscopies. Every device has its own focal length, aperture, and resolution. There are no large datasets with standardized parameters. 
Finally, polyp segmentation datasets are expensive because they depend on the annotations of qualified professionals with limited available time.

We propose an unsupervised method to detect polyps that does not require annotations by combining 3D rendering and a CycleGAN \cite{CycleGAN2017}. 
First, we produce artificial colons and polyps based on a set of parameters. Annotations of the location of the polyps are automatically generated by the 3D engine.
Second, the synthetic images are used alongside real images to train a CycleGAN. The CycleGAN is used to make the synthetic images appear more realistic.
Finally, we train a HarDNeT-based model \cite{chao2019hardnet}, a state-of-the-art polyp segmentation architecture, with the realistic synthetic data and our self-generated synthetic labels.

The contributions of this paper are as follows:
\begin{itemize}

\item
To the best of our knowledge, we are the first to train a polyp segmentation model with zero annotations from the real world.

\item 
We propose a pipeline that preserves the self-generated annotations when shifting the domain from synthetic to real.

\item 
We release Synth-Colon (see Figure \ref{dataset}), the largest synthetic dataset for polyp segmentation including additional data such as depth and 3D mesh.

\end{itemize}

The remainder of the paper is structured as follows: Section 2 reviews relevant work, Section 3 explains our method, Section 4 presents the Synth-Colon dataset, Section 5 describes our experiments, and Section 6 concludes the paper."

\section{Related work}
Here we briefly review some relevant works related to polyp segmentation and synthetic data.

\subsection{Polyp segmentation}

\begin{figure}
\centering
\includegraphics[width=70px]{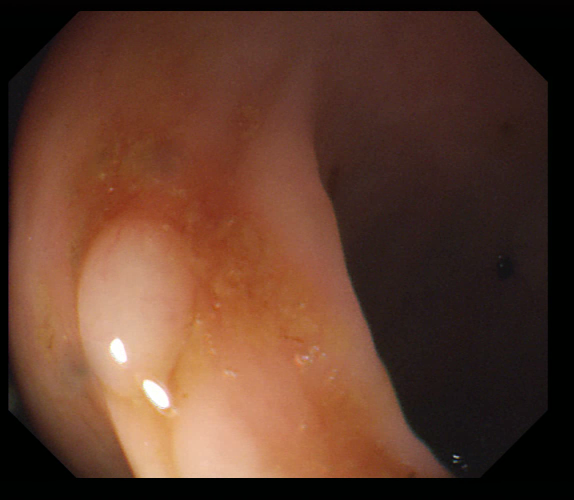}
\includegraphics[width=70px]{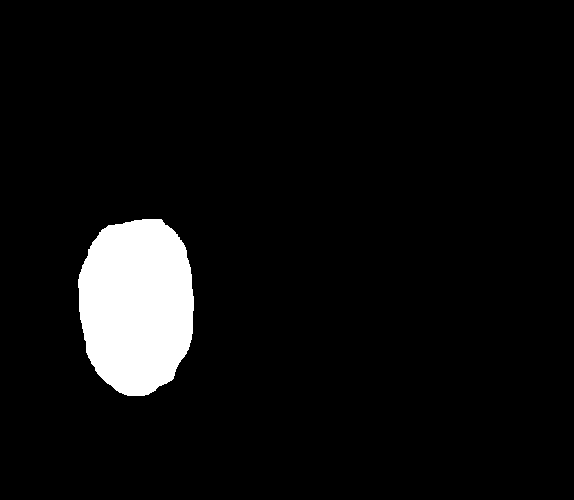}
\includegraphics[width=70px]{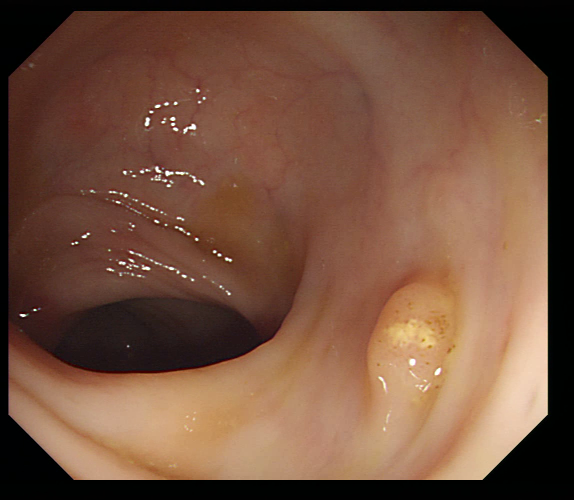}
\includegraphics[width=70px]{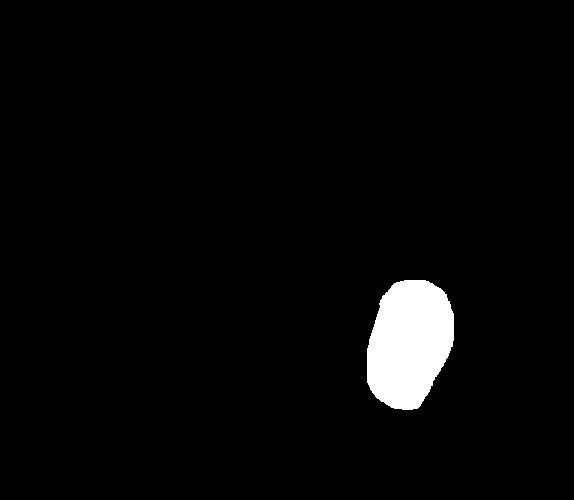}
\caption{Real samples from CVC-ColonDB with the corresponding annotation made by a medical professionals indicating the location of cancerous polyps.} \label{real_samples}
\end{figure}

Early polyp segmentation was based in the texture and shape of the polyps. For example, Hwang et al.~\cite{hwang2007polyp} used ellipse fitting techniques based on shape. However, some corectal polyps can be small (5mm) and are not detected by these techniques. In addition, the texture is easily confused with other tissues in the colon as can be seen in Figure \ref{real_samples}.

With the rise of convolutional neural networks (CNNs) \cite{lecun2015deep} the problem of the texture and shape of the polyps was solved and the accuracy was substantially increased. Several authors have applied deep convolutional networks to the polyp segmentation problem.
Brandao et al.~\cite{brandao2017fully} proposed to use a fully convolutional neural network based on the VGG \cite{simonyan2014very} architecture to identify and segment polyps. Unfortunately, the small datasets available and the large number of parameters make these large networks prone to overfitting. Zhou et al.~\cite{zhou2018unet++} used an encoder-decoder network with dense skip pathways between layers that prevented the vanishing gradient problem of VGG networks. They also significantly reduced the number of parameters, reducing the amount of overfitting. More recently, Chao et al.~\cite{chao2019hardnet} reduced the number of shortcut connections in the network to speed-up inference time, a critical issue when performing real-time colonoscopies in high-resolution. They focused on reducing the memory traffic to access intermediate features, reducing the latency. Finally, Huang et al.~\cite{huang2021hardnetmseg} improved the performance and inference time by combining HarDNet~\cite{chao2019hardnet} with a cascaded partial decoder~\cite{wu2019cascaded} that discards larger resolution features of shallower layers to reduce latency.

\subsection{Synthetic data}

The limitation of using large neural networks is that they often require large amounts of annotated data. This problem is particularly acute in medical imaging due to problems in privacy, standardization, and the lack of professional annotators. Table \ref{datasets} shows the limited size and resolution of the datasets used to train and evaluate existing polyp segmentation models.
The lack of large datasets for polyp segmentation can be addressed by generating synthetic data. 

Thambawita et al.~\cite{thambawita2021singan} used a generative adversarial network (GAN) to produce new colonoscopy images and annotations. They added a fourth channel to SinGAN \cite{rottshaham2019singan} to generate annotations that are consistent with the colon image. They then used style transfer to improve the realism of the textures. Their results are excellent considering the small quantity of real images and professional annotations that are used. Gao et al.~\cite{gao2020adaptive} used a CycleGAN to translate colonoscopy images to polyp masks. In their work, the generator learns how to segment polyps by trying to fool a discriminator.

\begin{table}[t]
\centering
\caption{Real polyp segmentation datasets size and resolution.}\label{datasets}
\begin{tabular}{lll}
\toprule
Dataset &  \#Images & Resolution\\
\midrule
CVC-T \cite{vazquez2017benchmark} & 912 & 574 x 500\\
CVC-ClinicDB \cite{bernal2015wm} &  612 & 384 x 288\\
CVC-ColonDB \cite{tajbakhsh2015automated} & 380 & 574 x 500\\
ETIS-LaribPolypDB \cite{silva2014toward} & 196 & 1225 x 966\\
Kvasir \cite{jha2020kvasir} & 1000 & Variable\\
\bottomrule
\end{tabular}
\end{table}

Synthetic images combined with generative networks have also been widely used in the depth prediction task \cite{mahmood2018unsupervised,rau2019implicit}. This task helps doctors to verify that all the surfaces in the colon have been analyzed. Synthetic data is essential for this task because of the difficulties to obtain depth information in a real colonoscopy.

Unlike previous works, our method is entirely unsupervised and does not require any human annotations. We automatically generate the annotations by defining the structure of the colon and polyps and transferring the location of the polyps to a 2D mask. The key difference between our approach and other state-of-the-art is that we combine 3D rendering and generative networks. First, the 3D engine defines the structure of the image and generates the annotations. Second, the adversarial network makes the images realistic.

Similar unsupervised methods have also been successfully applied in other domains like crowd counting. For example, Wang et al.~\cite{wang2019learning} render crowd images from a video game and then use a CycleGAN to increase the realism.

\section{Method}

Our approach is composed of three steps: first, we procedurally generate colon images and annotations using a 3D engine; second, we feed a CycleGAN with images from real colonoscopies and our synthetic images; finally, we use the realistic images created by CycleGAN to train an image segmentation model.

\subsection{3D colon generation}

\begin{figure}[t]
\centering
\includegraphics[width=200px]{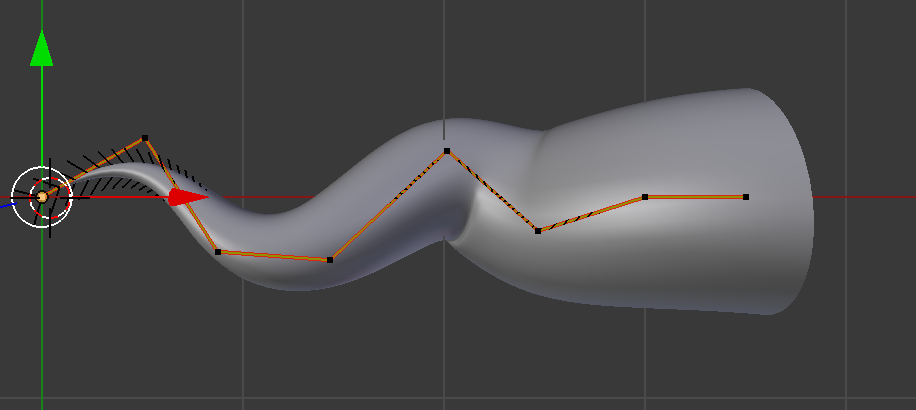}
\caption{The structure of the colon is composed by 7 segments to simulate the curvature of the intestinal tract.} \label{wireframe}
\end{figure}

The 3D colon and polyps are procedurally generated using Blender, a 3D engine that can be automated via scripting.

Our 3D colons structure is a cone composed by 2454 faces. Vertices are randomly displaced following a normal distribution in order to simulate the tissues in the colon. Additionally, the colon structure is modified by displacing 7 segments as in Figure \ref{wireframe}. For the textures we used a base color [0.80, 0.13, 0.18] (RGB). For each sample we shift the color to other tones of brown, orange and pink. One single polyp is used on every image, which is placed inside the colon. It can be either in the colon's walls or in the middle. Polyps are distorted spheres with 16384 faces. Samples with polyps occupying less than 20,000 pixels are removed.

\begin{figure}[t]
\centering
\includegraphics[width=70px]{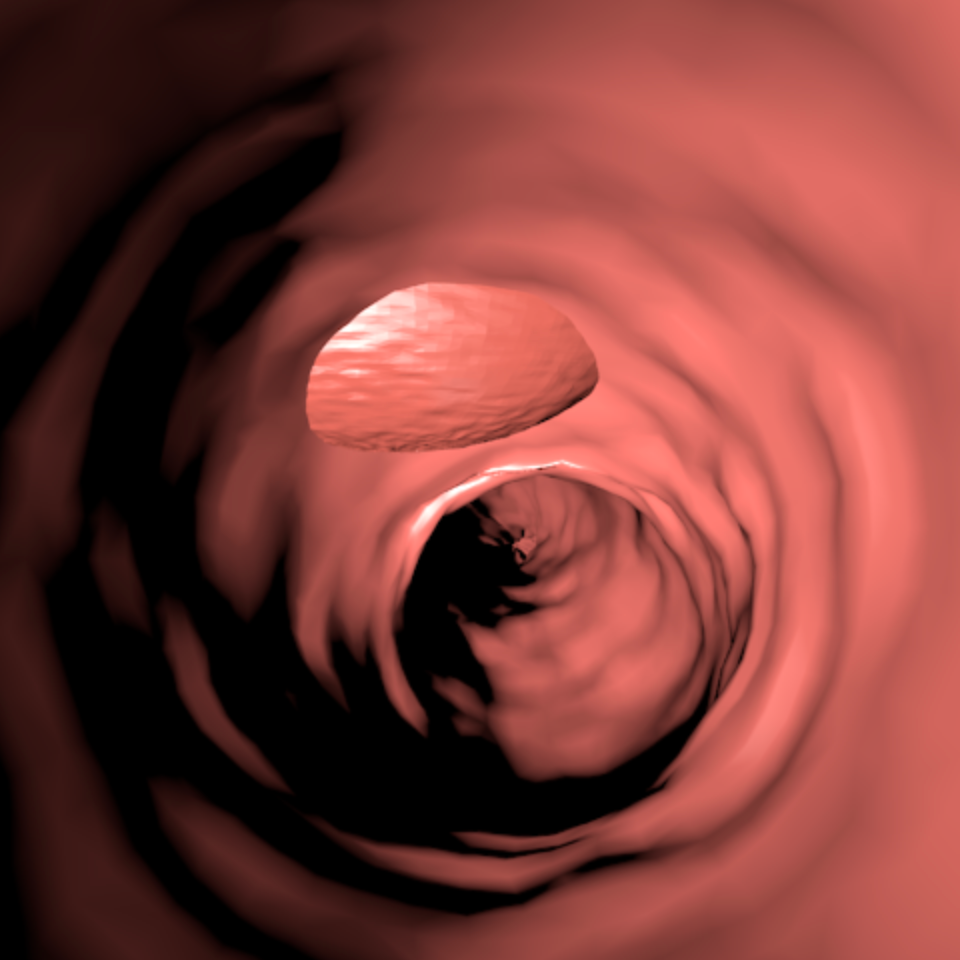}
\includegraphics[width=70px]{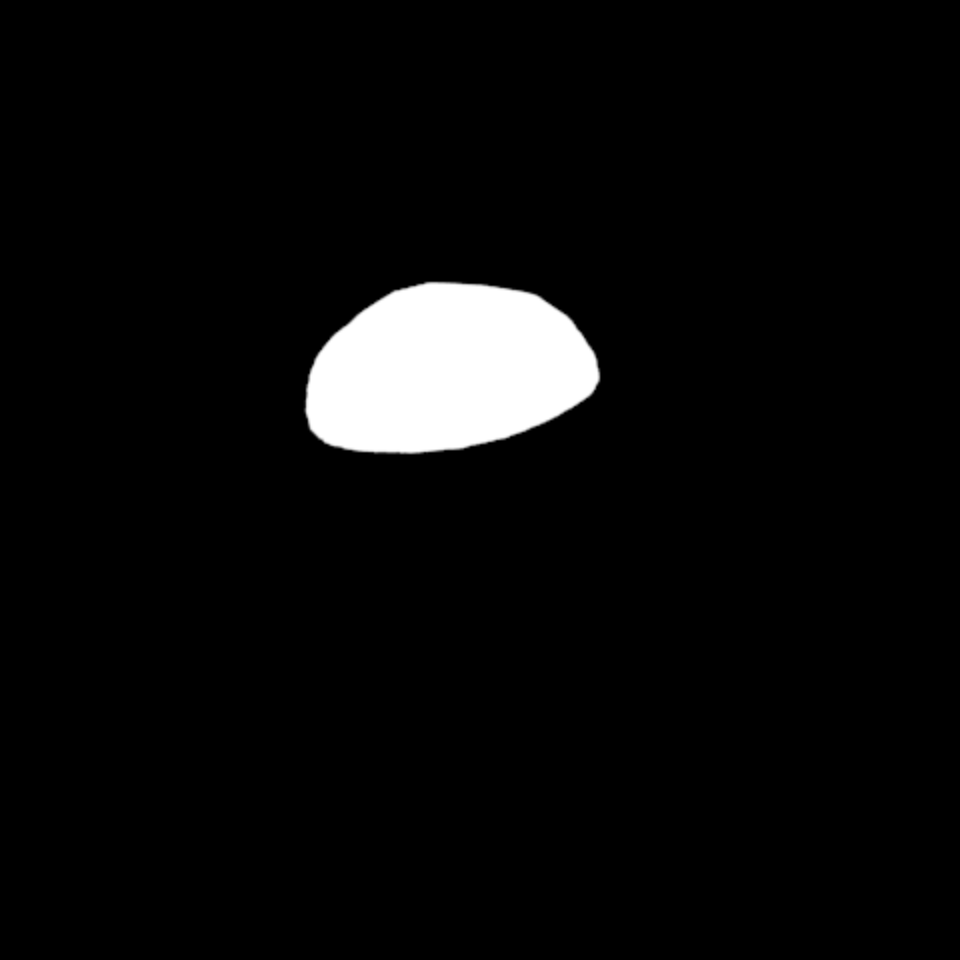}
\includegraphics[width=70px]{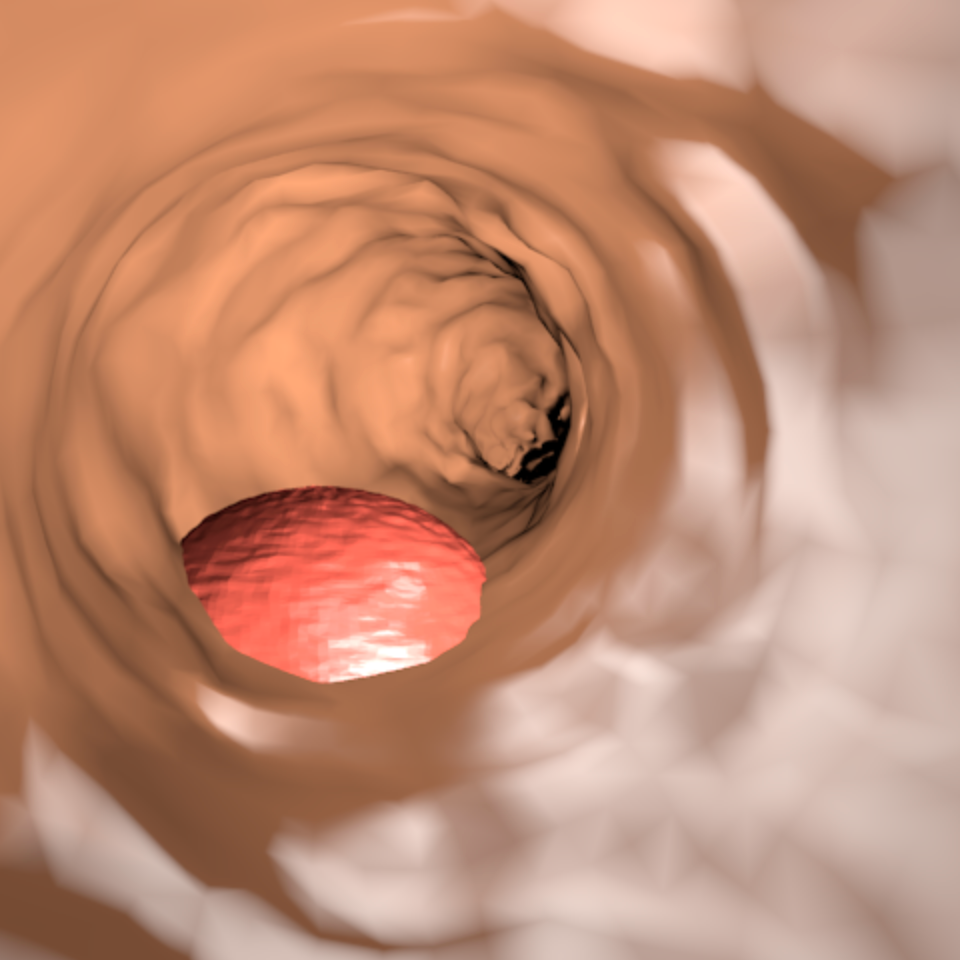}
\includegraphics[width=70px]{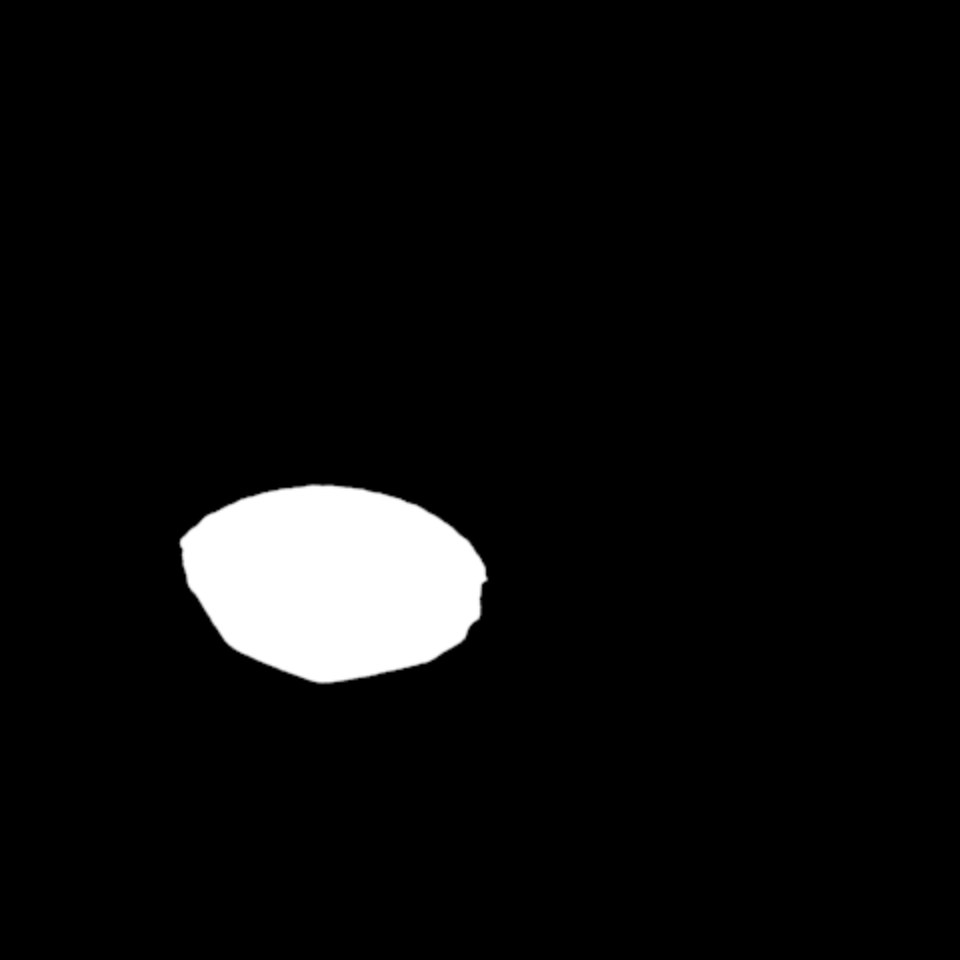}
\caption{Synthetic colons with corresponding annotations rendered using a 3D engine.} \label{synth_colons}
\end{figure}

Lighting is composed by a white ambient light, two white dynamic lights that project glare into the walls, and three negative lights that project black light at the end of the colon. We found that having a dark area at the end helps CycleGAN to understand the structure of the colon. The 3D scene must be similar to real colon images because otherwise, the CycleGAN will not translate properly the images to the real-world domain. Figure \ref{synth_colons} illustrates the images and ground truth generated by the 3D engine.

\subsection{CycleGAN}
\begin{figure}
\centering
\includegraphics[width=330px]{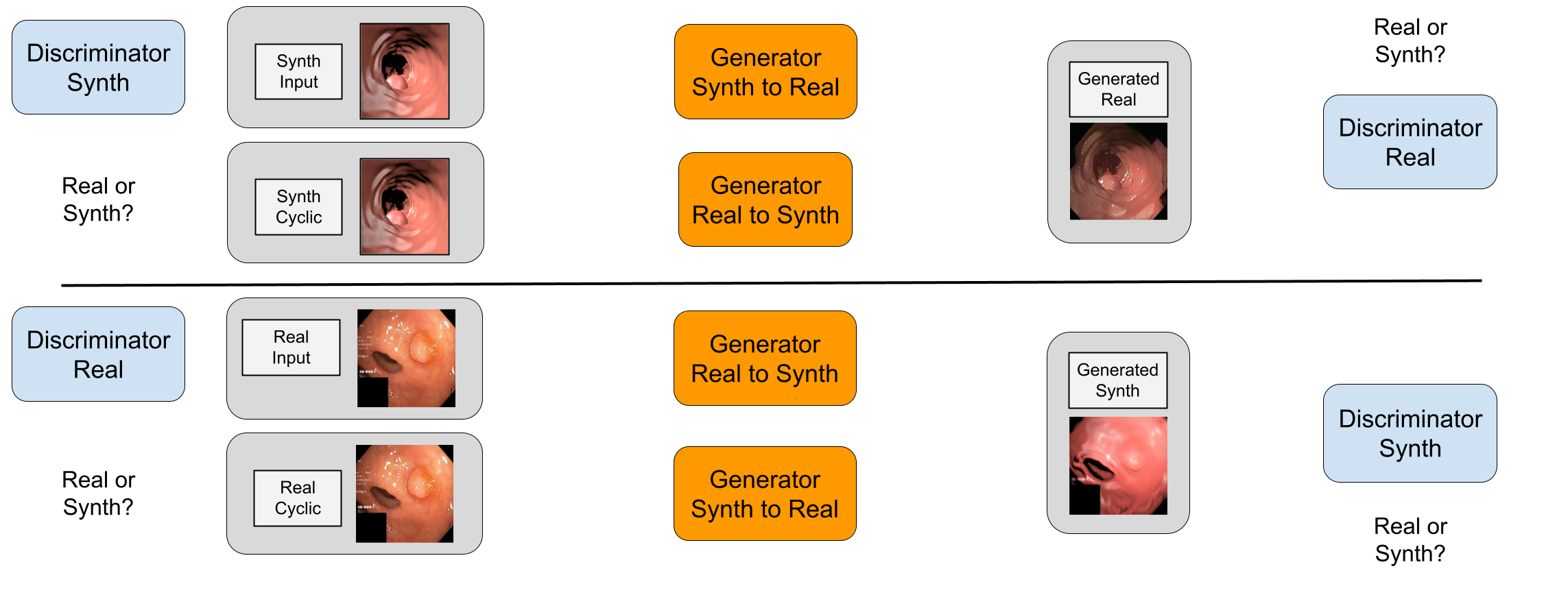}

\caption{Our CycleGAN architecture. We train two generator models that try to fool two discriminator models by changing the domain of the images.} \label{cyclegan_arch}
\end{figure}

A standard CycleGAN composed by two generators and two discriminators is trained using real images from colonoscopies and synthetic images generated using the 3D engine as depicted in Figure \ref{cyclegan_arch}. We train a CycleGAN for 200 epochs and then we infer real images in the ``Generator Synth to Real" model, producing realistic colon images.

\begin{figure}
\centering
\includegraphics[width=60px]{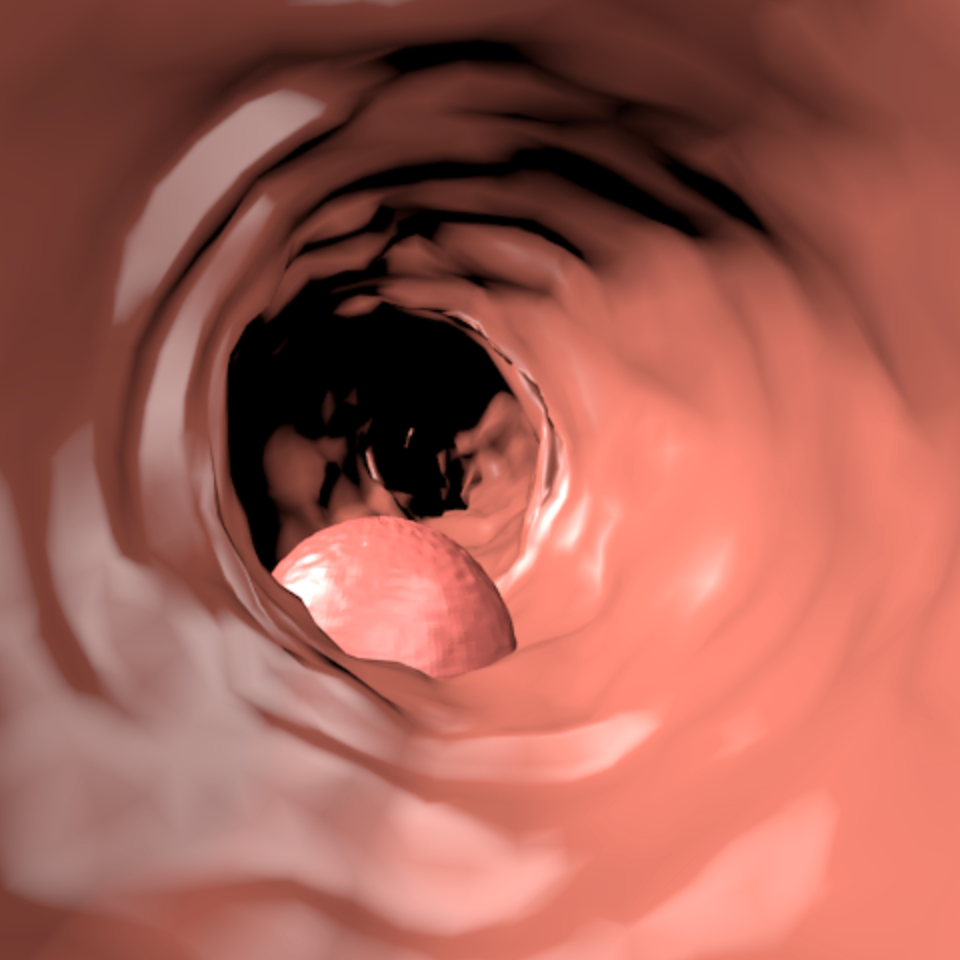}
\includegraphics[width=60px]{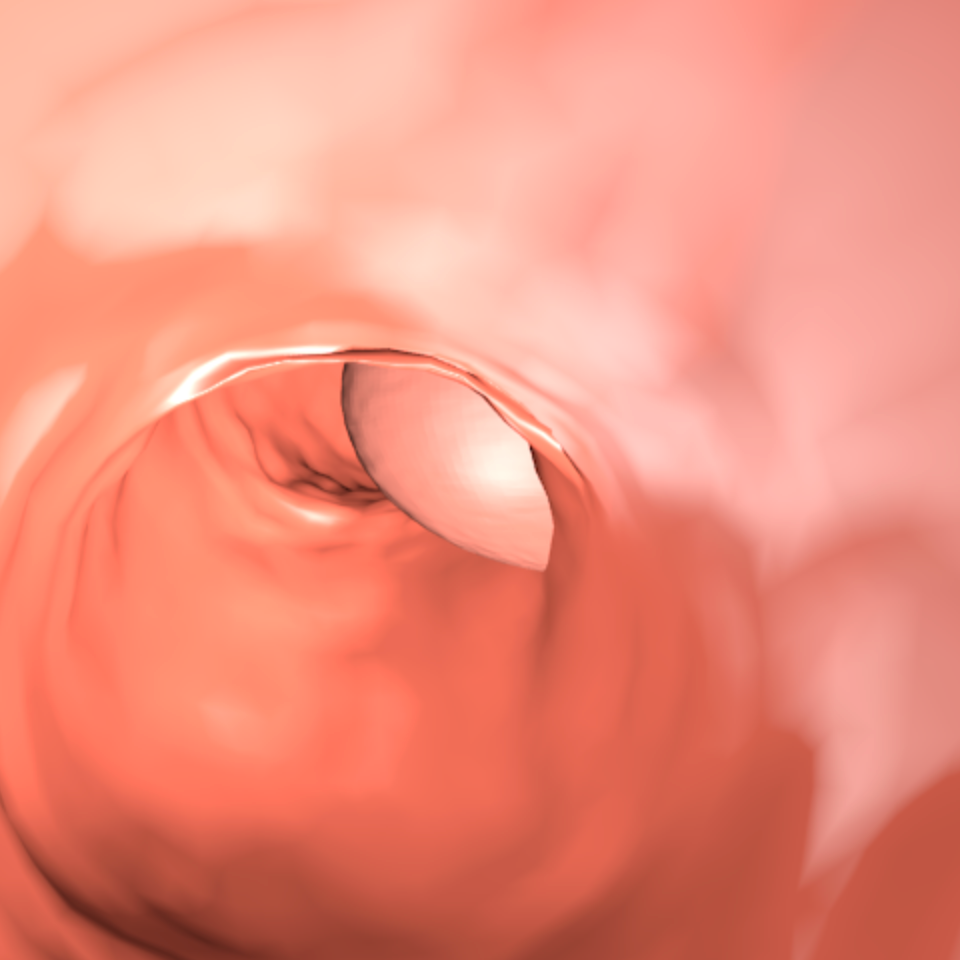}
\includegraphics[width=60px]{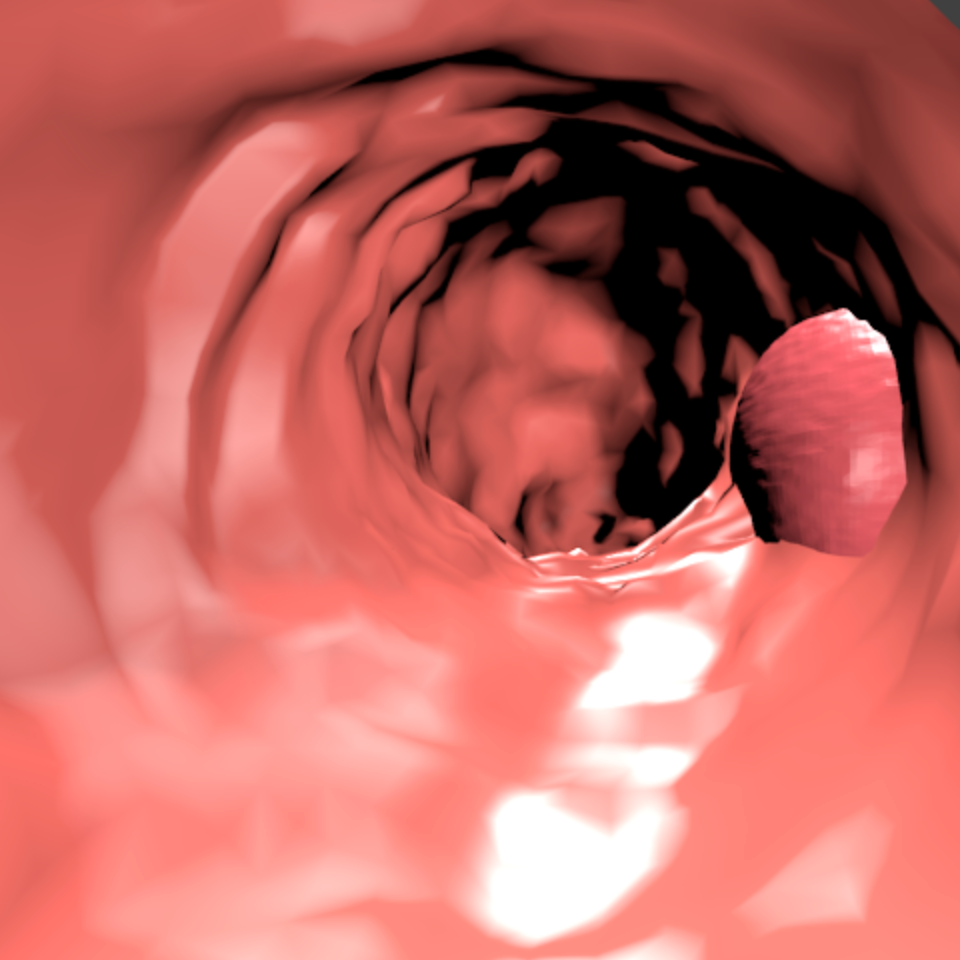}

\includegraphics[width=60px]{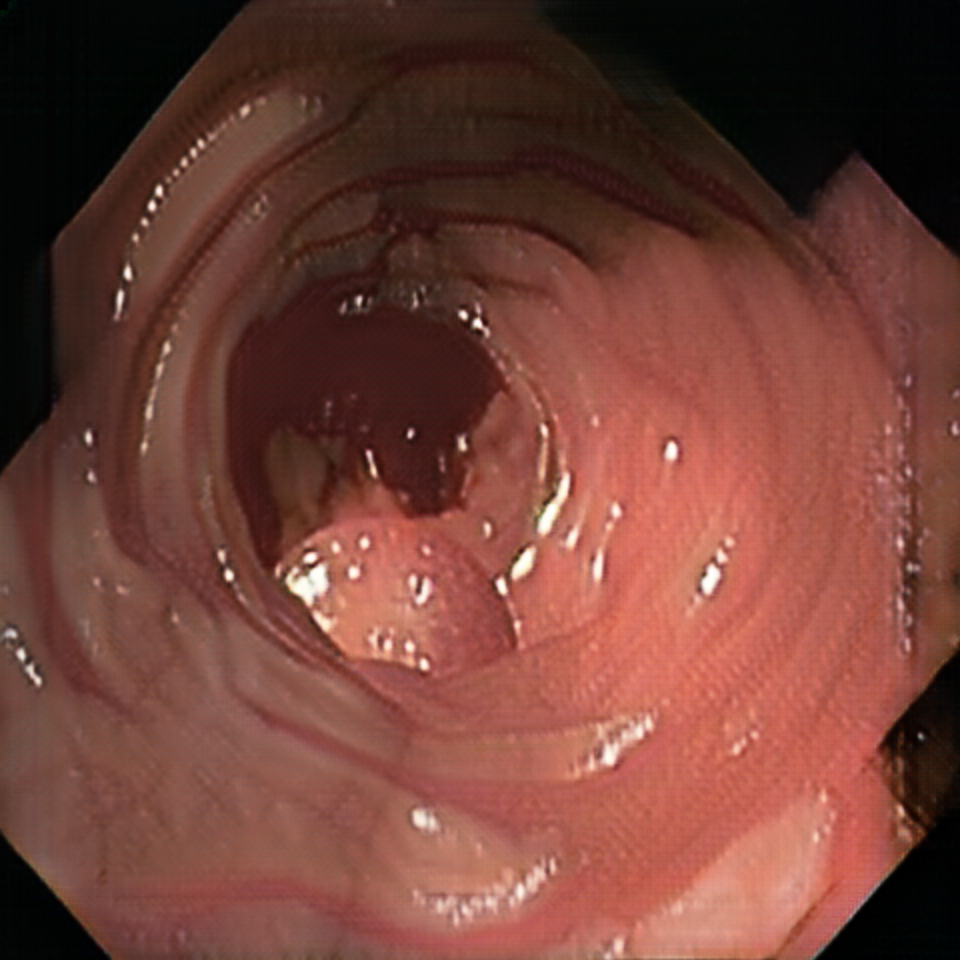}
\includegraphics[width=60px]{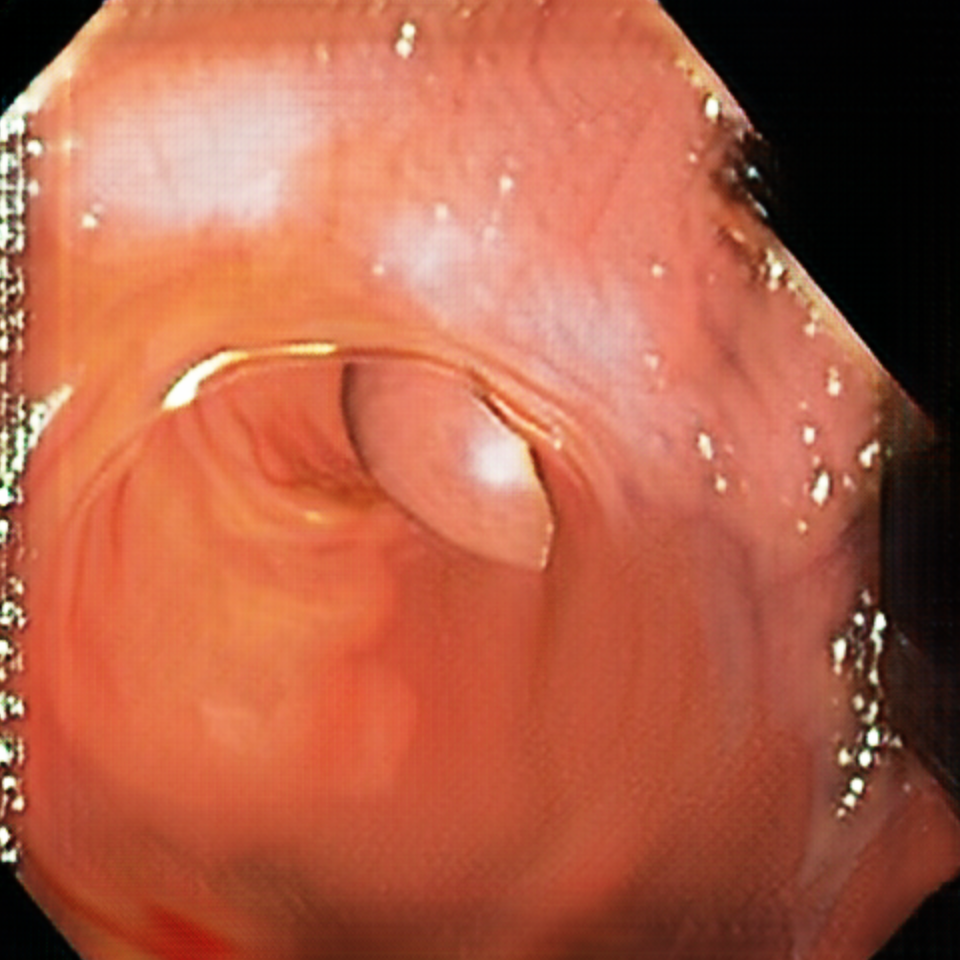}
\includegraphics[width=60px]{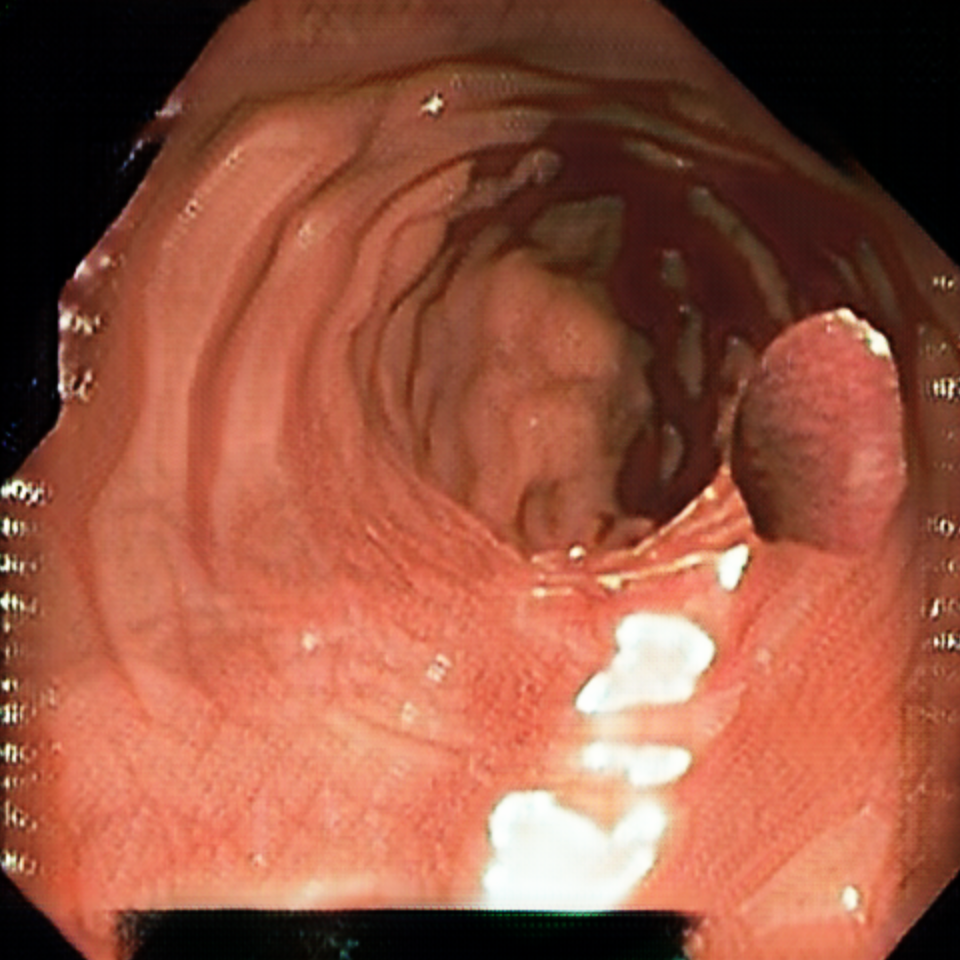}

\caption{Synthetic images (first row) and realistic images generated by our CycleGAN (second row).} \label{cyclegan_samples}
\end{figure}

Figure \ref{cyclegan_samples} displays synthetic images before and after the CycleGAN domain adaptation. Note that the position of the polyps is not altered. Hence, the ground truth information generated by the 3D engine is preserved.

\subsection{Polyp segmentation}
After creating a synthetic dataset that has been adapted to the real colon textures, we train an image segmentation model.
We used the HarDNeT-MSEG \cite{huang2021hardnetmseg} model architecture because of its real-time performance and high accuracy. We use the same hyperparameter configuration as in the original paper.

\section{Synth-Colon}
We publicly release Synth-Colon, a synthetic dataset for polyp segmentation. It is the first dataset generated using zero annotations from medical professionals. The dataset is composed of \numprint{20000} images with a resolution of 500$\times$500.  Synth-Colon additionally includes realistic colon images generated with our CycleGAN and the Kvasir training set images.
Synth-Colon can also be used for the colon depth estimation task \cite{rau2019implicit}  because we provide depth and 3D information for each image. Figure \ref{dataset} shows some examples from the dataset.
In summary, Synth-Colon includes:
\begin{itemize}

\item Synthetic images of the colon and one polyp.
\item Masks indicating the location of the polyp.
\item Realistic images of the colon and polyps. Generated using CycleGAN and the Kvasir dataset.
\item Depth images of the colon and polyp.
\item 3D meshes of the colon and polyp in OBJ format.

\end{itemize}

\section{Experiments}

\subsection{Metrics}

We use two common metrics for evaluation. The mean Dice score, given by:
\begin{equation}
    \mathrm{mDice} = \frac{2 \times tp}{2\times tp+fp+fn},
\end{equation}
and the mean intersection over union (IoU):
\begin{equation}
    \mathrm{mIoU} = \frac{tp}{tp+fp+fn},
\label{iou}
\end{equation}
where in both forumlae, $tp$ is the number of true positives, $fp$ the number of false positives, and $fn$ the number of false negatives.

% We used the mean Dice (\ref{dice}) and mean IoU (Intersection over Union) (\ref{iou}) for quantitative evaluation, similar to \cite{huang2021hardnetmseg}.

% \begin{equation}
% \frac{2 \times tp}{2\times tp+fp+fn}
% \label{dice}
% \end{equation}

% \begin{equation}
% \frac{tp}{tp+fp+fn}
% \label{iou}
% \end{equation}

\subsection{Evaluation on real polyp segmentation datasets}
We evaluate our approach on five real polyp segmentation datasets. Table \ref{unsupervised} shows the results obtained when training HarDNeT-MSEG \cite{huang2021hardnetmseg} using our synthetic data. Note that our method is not using any annotations.
Results are satisfactory considering the fact that labels have been generated automatically. We found that training the CycleGAN with only the images from the target dataset performs better than training the CycleGAN with all the datasets combined, indicating a domain gap among the real-world datasets.

\begin{table}[h]
\centering
\caption{Evaluation of our synthetic approach on real-world datasets. The metrics used are mean Dice similarity index (mDice) and mean Intersection over Union (mIoU).}\label{unsupervised}
\addtolength{\tabcolsep}{2pt}    

\resizebox{\textwidth}{!}{%
\begin{tabular}{l@{\qquad}cccccccccccc}
  \toprule
  &
  \multicolumn{2}{c}{CVC-T} & \multicolumn{2}{c}{ColonDB} & \multicolumn{2}{c}{ClinicDB} & \multicolumn{2}{c}{ETIS} & \multicolumn{2}{c}{Kvasir} \\
%   \cmidrule{2-12}
  & mDice & mIoU & mDice & mIoU & mDice & mIoU & mDice & mIoU & mDice & mIoU \\
  \midrule
  U-Net \cite{ronneberger2015u} & 0.710 & 0.627 & 0.512 & 0.444 & 0.823 & 0.755 & 0.398 & 0.335 & 0.818 & 0.746 \\
  SFA \cite{fang2019selective} & 0.467 & 0.329 & 0.469 & 0.347 & 0.700 & 0.607 & 0.297 & 0.217 & 0.723 & 0.611 \\
  PraNet \cite{fan2020pranet} & 0.871 & 0.797 & 0.709 & 0.640 & 0.899 & 0.849 & 0.628 & 0.567 & 0.898 & 0.840 \\
  HarDNet-MSEG \cite{huang2021hardnetmseg} & 0.887 & 0.821 & 0.731 & 0.660 & 0.932 & 0.882 & 0.677 & 0.613 & 0.912 & 0.857 \\
  Synth-Colon (ours) & 0.703 & 0.635 & 0.521 & 0.452 & 0.551 & 0.475 & 0.257 & 0.214 & 0.759 & 0.527 \\
  \bottomrule
\end{tabular}}

% \begin{tabular}{lll}

% \toprule
% Dataset &  mDice & mIoU\\
% \midrule
% CVC-T \cite{vazquez2017benchmark} & 0.703 & 0.635\\
% CVC-ColonDB \cite{tajbakhsh2015automated} & 0.521 & 0.452\\
% CVC-ClinicDB \cite{bernal2015wm} &  0.551 & 0.475\\
% ETIS-LaribPolypDB \cite{silva2014toward} & 0.257 & 0.214\\
% Kvasir \cite{jha2020kvasir} & 0.759 & 0.527\\
% \bottomrule
% \end{tabular}

\end{table}

% [Image of our predicted masks vs ground truth]
% [Failure case of our masks prediction]

\subsection{Study with limited real data}

In this section we evaluate how our approach based on synthetic imagery and domain adaptation compares with the fully supervised state-of-the-art HarDNeT-MSEG network when there are fewer training examples available. We train the CycleGAN used in the proposed approach, without ground truth segmentation labels, on progressively larger sets of imagery, and compare this with the supervised method trained on the same amount of labelled imagery. Table \ref{mix} shows the results of the experiment, which demonstrates that synthetic data is extremely useful for domains where annotations are very scarce. While our CycleGAN can produce realistic images with a small sample of only five real images, supervised methods require many images and annotations to achieve good performance. Table \ref{mix} shows that our unsupervised approach is useful when there are less than 50 real images and annotations. Note that zero images here means there is no domain adaptation via the CycleGAN.

\begin{table}
\centering
\caption{Evaluation of the proposed approach on the Kvasir dataset when few real images are available. The performance is measured using the mean Dice metric.}\label{mix}
\begin{tabular}{l@{\hskip 3em}l@{\hskip 1em}l}
\toprule
 & Synth-Colon (ours) & HarDNeT-MSEG \cite{huang2021hardnetmseg}\\
\midrule
0 images & 0.356 & -\\
5 images & 0.642 & 0.361\\
10 images & 0.681 & 0.512\\
25 images & 0.721 & 0.718\\
50 images & 0.735 & 0.781\\
900 (all) images & 0.759 & 0.912\\
\bottomrule
\end{tabular}
\end{table}

\section{Conclusions}
We successfully trained a polyp segmentation model without annotations from doctors. We used 3D rendering to generate the structure of the colon and generative adversarial networks to make the images realistic, and demonstrated that it can perform quite reasonably in several datasets, even outperforming some fully supervised methods in some cases.
We hope this study can help aligning synthetic data and medical imaging in future. As future work, we will explore how to include our synthetic annotations in the CycleGAN.

% \begin{figure}
% \centering
% \includegraphics[width=70px]{images/real_to_synth_1r.png}
% \includegraphics[width=70px]{images/real_to_synth_2r.png}

% \includegraphics[width=70px]{images/real_to_synth_1s.png}
% \includegraphics[width=70px]{images/real_to_synth_2s.png}
% \caption{Real samples (first row) transfered to the synthetic domain (second row) using CycleGAN.} \label{real_to_synth}
% \end{figure}

\bibliographystyle{splncs04}
\bibliography{citations}

\end{document}